\documentclass[12pt]{article}
\usepackage[english]{babel}
\usepackage{amsmath,enumerate}
\usepackage{a4wide}
\usepackage{hyperref}

\newcommand{\nobracket}{}
\newcommand{\tmem}[1]{{\em #1\/}}
\newcommand{\tmmathbf}[1]{\ensuremath{\boldsymbol{#1}}}
\newcommand{\tmop}[1]{\ensuremath{\operatorname{#1}}}
\newcommand{\tmstrong}[1]{\textbf{#1}}
\newcommand{\tmtextbf}[1]{\text{{\bfseries{#1}}}}
\newcommand{\tmtextit}[1]{\text{{\itshape{#1}}}}
\newenvironment{enumeratealpha}{\begin{enumerate}[a{\textup{)}}] }{\end{enumerate}}
\newenvironment{enumeratenumeric}{\begin{enumerate}[1.] }{\end{enumerate}}
\newenvironment{enumerateroman}{\begin{enumerate}[i.] }{\end{enumerate}}
\newenvironment{enumerateromancap}{\begin{enumerate}[I.] }{\end{enumerate}}
\newenvironment{itemizeminus}{\begin{itemize} }{\end{itemize}}
\newenvironment{quoteenv}{\begin{quote} }{\end{quote}}

\begin{document}

\title{The seven laws of Quantum Mechanics : banishing the bogeys}

\author{Urjit A. Yajnik\footnote{Email : yajnik@iitb.ac.in}, \\
{\tmem{Physics Department, Indian Institute of Technology Bombay, Mumbai 400076}}}
\date{}
\maketitle

\begin{abstract}
  \tmtextbf{}The laws of quantum mechanics are couched in subtle mathematical language. The laws are not usually stated in a compact pedagogical form. Here I present a possible way to correct this. Essential facts can be distilled into seven statements that are easy to remember and easily referred back. Also, the current teaching of   quantum mechanics is laden with words of negative connotations, originating as they did during the early decades of the subject when the subject was intellectually still puzzling. A wide variety of experiments in the intervening decades, not least those that were awarded the Nobel Prize of 2022 amply affirm the validity and substantial ``reality'' of Quantum Mechanics as a theory. I take a few of the inadequacies of classical framework to illustrate that some of the complaints against Quantum Mechanics are patently misplaced. Finally I discuss the bogeys such as ``wave article duality'', ``uncertainty'', ``indistinguishability'' ``statistics'' and ``entanglement'' and advocate adopting better terminology to save new learners from the old biases.
\end{abstract}

\tableofcontents

\section{Why number the laws?}\label{sec:whynumber}

We teach three laws of Newton, three laws of Thermodynamics and then include a
zeroth law as well, Four Maxwell's equations of \ Electromagnetism etc. This
simple mnemonic pedagogical device is missing from the teaching of Quantum
Mechanics. Stating them as a numbered set of laws would make it easier to
remember them. This can also be used to set \ logical primacy and separation
between them. Then the critical comments would be easier to make, as in
discussing the subtlety of ``Newton's third law''.

Clearly differentiated laws also expose possible pitfalls. \ For example
Newton's first law is essentially due to Galileo, and after Newton's second
law, the former can be taken to correspond to the special case of no applied 
force. Is it logically independent? Or does it still have a logical primacy due to direct
observability? Furthermore, substantial discussion is needed to explicate whether 
the second law defines mass or defines force\cite{KibbleClassMech}. Thus, setting apart the First Law asserts
its validity independent of the dynamical framework of Newton. We hope that in
the following discussion on Quantum Mechanics we are able to place logically 
independent facts and
rules under independent laws. After doing that we will be able to identify the
source of various concerns people have about Quantum Mechanics. It is also
hoped that such laws would be useful to new learners. Considering the strides
quantum science is making, we may soon be teaching such laws in the high
school. \ Clear laws stated without the negativity imposed by the older
generation would reduce the hesitation and suspicion in accepting and applying
the laws. Until some completely new phenomenon demanding a revision comes to the
fore we can be at peace with the laws of Quantum Mechanics, now known for a
century.

This article is mostly motivated by Dirac's elegant exposition in his famous
textbook\cite{DiracPQM}. This book is most widely praised but least widely read. Teachers and
even experts are heard saying ``.. but that is a difficult book to read''. It
happens that Dirac's exposition is mathematical, but its style is physics. It
makes physicists feel it is too abstract while the mathematicians need more
sophisticated underpinnings. However the book is really precious for its
clarity and elegance. Among many other points of great clarity in this book
is the introduction to many-body quantum mechanics, laying bare the mistaken
origins of the term \ ``second quantisation''.

\subsection{Welcoming the new principle}\label{sec:newprinciple}

To be specific, our motivation in what follows is to propose that the language
of the ``wave function'' has been the slow poison of quantum mechanics. At its
inception in 1926 this language was the easiest to follow for physicists.
Further, it serves a very good tool for visualising electronic orbitals and
shall remain a very useful language for many settings. However its overuse,
and debating many points of principle in that language, has lead to
misconceptions and a feeling of incompleteness of the subject and inadequacy
of the framework.

The facts look rather different if we grasp Heisenberg's seminal contribution.
Heisenberg's 1925 paper \cite{Heisenber1925} called for complete refurbishing of mechanics :
\begin{quotation}
``... one realises that ... even for the simplest quantum-theoretical problems the validity of classical mechanics simply cannot be maintained. In this situation it seems sensible to discard all hope of observing hitherto unobservable quantities such as the position and the period of the electron, ... Instead it seems more reasonable to try to establish theoretical quantum mechanics ... in which only relations between observable quantities occur.''
\end{quotation}
According to this point of view the notion of classical trajectory $X(t)$ needs to be abandoned. What can be really ascertained physically is only
the transition amplitude $X_{ab}$ for a particle to have been once seen at $a$ and then at $b$. We must thus forgo the knowledge of what may have
happened in between as  inaccessible {\tmem{in principle}}.
What does
not exist does not need any organising principle. No ontology no epistemology.

Heisenberg replaces the trajectory by an array of numbers $X_{ab}$ and
additionally $\dot{X}_{ab}$ for the velocity \cite{Aitchison:2004cic}. He then proceeds
to check the mutual commutability of these arrays of numbers. This is what we
call matrix mechanics now. It was Dirac's insight\cite{Gottfried:2010kn} to note that instead of
velocity one should use the canonically conjugate $P_{ab}$ and when this is
done we get the very elegant analogy of the commutation relations (CR) needed
in Quantum Mechanics to the Poisson brackets (PB) of Classical Mechanics.
However, unlike the PB which are based on real analysis, the quantum
mechanical brackets CR imply that the dynamical variables cannot be
represented by mere numbers, but need matrices for their representation.

We may immediately feel very anxious about this radical framework to be used
for something as obvious as a trajectory. But we shall return in a later
section to argue that the real culprit is the Newtonian notion of
instantaneous velocity, which is at least as ``unreasonable'' as these quantum
hypotheses. However for the moment let us consider the two positive thoughts
put across by Dirac. The existence of $\hbar$  is the natural
demarcation between the ``macroscopic'' and the ``microscopic''. It sets the
scale below which the microscopic world begins. Otherwise the world should
remain self-similar under infinite subdivision. But Dalton, Avogadro, Boltzmann, Thomson,
Rutherford and others had already investigated that the
microscopic world is atomistic and rather different.

Dirac's second point is that quantum principles need not be viewed negatively,
as a loss of familiar concepts. Rather, there is a new positive principle as a
compensation so to speak. This is the principle of linear superposition.
Agreed, there is an added layer of abstraction. ``State of a system'' in
quantum mechanics is not given by a list of values that are guaranteed to be
the outcome of measurement. However, this abstract state vector does obey the
Principle of Superposition.

The uncanny nature of the situation is best brought home to students by
the example of the position and velocity of a ball in a playing field. 
Common sense says that the possible states of the ball 
are given by positions $\vec{R}_A$
and corresponding velocities $\vec{v}_A$ where $A$ denotes the point in space and 
$\vec{R}_A$ its position vector. But in quantum mechanics we get new 
valid ``states''  of the ball which are linear ``sum''s of states at $A_1$, 
$A_2$, ...$A_n$. The possible number of states increases manifolds, 
as the relative weightages vary from $0$ to $1$ in magnitude, can have relative 
complex phases, and can include any number $n$ of classical states. 
Thus the Quantum World offers multitudes of possibilities far 
beyond the classical imagination. Yet this is not a disaster. Dirac exhorts us 
to admire the simplicity  of the linear superposition principle rather 
than be baffled by it.

Thus the basic attitudes to quantum mechanics need to be changed. Presenting
the set of arguments that substantiate this appeal is the goal of this
article. \ In the following in Sec. \ref{sec:experiments} we 
summarise the classic
experiments that signal the novel behaviour, Sec. \ref{sec:postulates} 
presents the statement of the laws, not in
truly pedagogical form but as notes for knowledgeable peers, Sec. \ref{sec:classical} 
contains a discussion of a few key classical conceptions, where we 
argue that the latter are, to wit, as technical as the ones involved 
in quantum mechanics and indeed not in accord with observed reality. 
In Sec. \ref{sec:bogeys} I comment on the prevalent terminology, identifying 
the bogeys that need to be banished. Sec. \ref{sec:conclusion} contains the
concluding remarks.

\section{Characteristic experiments}\label{sec:experiments}

It will be important to distinguish between what is truly novel in the quantum
phenomena themselves, versus what makes us uneasy about the mathematical
framework. We shall try to show that whatever is novel is indeed rooted in the
phenomena themselves. More interestingly, as argued later, whatever makes us
uneasy about the formalism is perhaps no worse than the state of affairs in
the classical framework. \

The enigmas of quantum systems can be summarised as
\begin{enumeratealpha}
  \item The observer is free to choose what to measure, though the choices are limited in any one attempt at measurement.
  
  \item The system will produce probabilistic outcomes.
  
  \item Subsequently the system either ceases to exist completely, or the
  measured attribute becomes predictable with certainty.
\end{enumeratealpha}
Note the most uncanny feature of the last statement. In macroscopic world we
only see systems change or transform or redistribute. But the quantum world
allows entities to vanish forever, like a photon absorbed by an atom or the
neutrino by inverse beta decay in nuclei. Then the case of measured attributes becoming certain in values may  be understood as one form of extinction of the other values
of the attribute.

These points can be stated succinctly as

\begin{quoteenv}
  a') Subjectivity in the choice of measurement,
  
  b') probabilistic outcomes,
  
  c') objectivity of the post measurement state
\end{quoteenv}

We now summarise the well known phenomena which we may keep in mind as
intrinsic quantum behaviour
\begin{enumerate}
  \item {\tmstrong{Complementarity of description }}- The Davisson-Germer
  experiment idealised as double slit experiment.
  
  The electrons although particles can get redistributed to give the
  interference pattern displayed by wave phenomena. Extensive work has shown
  that one can recover particle like properties or wave like properties but
  not both in the same measurement.
  
  \item {\tmstrong{Quantised values in case of some observables}} - The
  Stern-Gerlach experiment
  
  A beam of polarised Silver atoms passes through a region of magnetic field
  pointing along $z$-axis. The emerging beam is split into precisely two
  streams. The only explanation is that the component of intrinsic spin of the
  Silver atoms along the applied field can have only two possible values.
  Further only one of these precise values can emerge in any measurement, and
  not an averaged value.
  
  \item {\tmstrong{Probabilistic outcomes}}
  
  Outcomes of measurements on identically prepared systems are not identical.
  We can at best associate a probability to any outcome. If the Stern-Gerlach
  type experiment is performed with a rarefied beam of atoms so that only one
  atom is passing the magnetic field region, we cannot predict which of the
  two orientations the atom will finally emerge in, only relative
  probabilities for the two outcomes.
  
  \item {\tmstrong{Indeterminate evolution during measurement process}}
  
  Related to the previous point is the independent fact that there is no
  theory for how the quantum state evolves ``during'' the measurement. The
  imprint of the quantum system is recorded and the quantum nature of that
  attribute then terminates. However there is no theoretical framework for
  describing the evolution of a quantum system into a residual
  system with specific value as it leaves its imprint in an apparatus.
  
  More generally the system itself may disappear such as charged particles in
  a Geiger counter. The result is a macroscopic current, and we have a quantum
  theory of how the single particle cascades into a current, but not a
  deterministic theory.
  
  \item {\tmstrong{Non-locality of states}}
  
  The quantum state can be spread over a macroscopic region. A state of two
  photons can be stretched over many meters. But as soon as the attributes of
  one of the photons are measured, the state evolves through that
  indeterminate evolution and the attributes of the other photon are
  instantaneously determined. \ One version of it is the famous EPR paradox.
However unreasonable classically, the validity of this outcome has finally 
been accepted and recognised by the Nobel Prize of 2022\cite{Nobel2022}.
  
  The non-locality is also manifested over time. We might suspect that once we
  have decided on which attribute to measure, the two photons mutually encode
  which one is to manifest which value before they are far apart. But since
  there is complementarity of which attribute to measure, we need not set up
  which attribute we wish to measure until the photons are really far apart.
  This is called ``delayed choice''. However even under delayed choice, the
  effect of measuring an attribute on one photon immediately determines the
  outcome of that attribute on the other photon. 
  
  \item {\tmstrong{Bose condensation and Pauli Exclusion Principle}}
  
  Perhaps the most radical departure from classical conceptions arises in the
  very notion of identity of the basic entities, particles, or more correctly,
  quanta. There are two fundamental aspects to this. One is that there are
  identical particles : the primary units are endowed with very few
  attributes, and all quanta of a particular species have just a few possibilities for
  these attribute. For example all electrons have exactly the same values of
  charge and mass, and can have one of two values for the projection of their
  spin along any measured direction. This is not encountered classically even in what
  we think are identical objects, say balls of same size, color and shine of polish. 
  Improving the precision of measurement always reveals the differences, often 
  taking continuum values.
  
  Further the quanta obey peculiar rules for collective states. Atomic physics
  verifies most directly that two electrons cannot occupy the exact same state
  of energy and spin. Likewise, the spectrum of ``light gas'' under ideal
  conditions of Black Body obeys Planck spectrum, which can be understood only
  if photons occupy their available energy states according to the rules
  discovered by Planck, Einstein and Bose.
  
  In summary there are identical particles and they obey strange rules for
  their collective states. These facts are foundational to the quantum world
  and the laws pertaining to them need to be formalised into the core of
  Quantum Mechanics. The law should not be postponed as auxiliary rules to be
  learnt in more advanced courses.
\end{enumerate}
\section{The postulates}\label{sec:postulates}

In this section we enunciate the laws in simple naive language. These are
stated more as pointers to what all practitioners of Quantum Mechanics are
quite familiar with. The idea is to lay out these rules in an order from the
more basic or elementary, moving towards those that build the structure
further. We skirt several subtleties about the Hilbert space and precise
meaning and varieties of measurement at the level of this presentations.
Stating the more detailed version will however not require a change to this
basic list.

Let us make a brief qualitative statement about what is at stake, {\'a} la
Dirac. We have a quantum system and we have some apparatus that will record
various clicks and ticks. The requirement is to set up a mathematical
framework that will make predictions about what clicks and ticks can result as
outcomes. The dynamical quantities that can be thus measured will be called
observables. Examples are charge, mass, spin projections, binding energies,
lifetimes, \ etc. This list needs to be established empirically for every new
quantum system one encounters.

The program of quantum mechanics is to identify a core set of variables which
should satisfy relatively simple kinematic conditions, the CR's, and all the
other observables should be expressible in terms of them. Fortuitously, but
with no guarantee, this set happens to be the same as the set of classical
canonical variables, satisfying the CR's that are, upto the fundamental unit
$\hbar$ the same as the classical PB's. Indeed all the observables, all the
relevant symmetry operators, and the dynamical evolution operator, can be
constructed out of this canonical set. This is not at all obvious, and indeed
it fails in a few major exceptional cases such as spin which has no canonical
representation. A failure also shows up in the occurrence of anomalies in 
advanced implementation of quantum principles, such as in Quantum Field Theory
and String Theory. But
majority of quantum observables do have simple classical analogy and that is
what has greatly facilitated the prediction and control of the quantum world.

The rules stated below essentially address the logical structure of these
constructs and the mathematics required to implement them. They are stated in
the simplest context of a single variable and a single observable etc, to keep
the statements compact. 

We shall use the convention
\[ \begin{array}{ll}
     | \Psi \rangle \nobracket \nobracket, |\varphi \rangle  &
     \text{A generic state}\\
     | \Psi t \rangle \nobracket \nobracket, |\varphi t \rangle & \text{A generic state displaying its time dependence} \\
     | x \rangle \nobracket \nobracket, | p \rangle \nobracket \nobracket 
     \hspace*{\fill}  & \text{A basis state labeled by a canonical
     variable}\\
     | \alpha \rangle \nobracket \nobracket, | l, m \rangle \nobracket \nobracket,
     | n \rangle \nobracket \nobracket & \text{Eigenstates of general
     observables labeled by their}\\
     & \text{eigenvalues}
   \end{array} \]
We are dispensing with the convention of putting a hat or a caret above an
operator, so long as there is no ambiguity.
\begin{enumerateromancap}
  \item {\tmstrong{State functions constitute a Hilbert space.}}
  \label{pos:Hilbert}
  
  The states of a quantum system obey linear superposition principle and have the structure of a
  complex vector space. Further,  for physical interpretation we need to endow this space with a hermitian
  inner product.
  \[ \langle \psi | \varphi \rangle = \langle \varphi | \psi \rangle^{\ast}
  \]
  The framework of Hilbert spaces is applicable with some caveats.
  
  \item {\tmstrong{Observables are realised as Hermitian
  operators.\label{pos:Observable} }}
  
  For an operator $A$, the hermitian conjugate or adjoint operator is defined
  as :
  \[ \langle A \psi | \varphi     \rangle \equiv \langle \psi | A^{\dag} \varphi \rangle  \]
  Hermitian operators are self-adjoint. All observables are represented by
  hermitian operators. Their eigenvalues will be the possible list of answers
  we get upon observation. The eigenstates corresponding to these eigenvalues
  can be used to construct a basis for the Hilbert space. There can be several
  independent choices of bases.
    
  \item {\tmstrong{Change of basis is implemented by Unitary
  operators}}\label{pos:Changebasis}
  
  If we change from a basis constructed using an observable $n$ to a basis
  constructed using an observable $\alpha$, then the change is implemented by
  a unitary transformation
  \[ | \alpha \rangle = \sum_n U_{\alpha n} | n \rangle \]
  \[ \sum_{\beta} (U^{\dag})_{m \beta} U_{\beta n} = \delta_{m, n} \qquad 
     \sum_m U_{\alpha m} (U^{\dag})_{m \beta} = \delta_{\alpha, \beta} \qquad
  \]
  where the Kronecker deltas need to be replaced by Dirac delta functions for
  continuum eigenvalues.
  
  \item {\tmstrong{Observations and probabilities are described by projection
  operators.\label{pos:projection}}}
  
  In an observation process we can only predict the probability for the system to
  emerge in a particular eigenvalue of the relevant observable. If the
  observable is $\alpha$ and the initial normalised state vector is
  represented in the $| \alpha_i \rangle \nobracket$ basis as
  \[ | \Psi \rangle = \sum_i C_i | \alpha_i \rangle, \]
  then the probability of getting the outcome $\alpha_r$ is $| C_r |^2$. 
  In Hilbert space this amounts to the state being subjected to a projection operator.
  
  This leads to a simple rule about the average measured value of that observable 
  under repeated measurement of identically prepared systems. 
  \[
  \langle A \rangle \equiv  \langle \Psi \vert A \vert \Psi \rangle \equiv  \langle \Psi \vert A \Psi \rangle =  \sum_i | C_i|^2  \alpha_i
  \]
  where the second equivalence connects a Physics convention with Hilbert space operation.
  
  \item {\tmstrong{Quantum kinematics.\label{pos:kinematics}}}
  
  The observables and other dynamical quantities introduced are operators and
  need not commute. With a sufficiently exhaustive set of dynamical
  quantities $\mathcal{O}_i$ we find that they satisfy a symplectic algebra
  \[ [\mathcal{O}_i, \mathcal{O}_j] \equiv \mathcal{O}_i \mathcal{O}_j
     -\mathcal{O}_j \mathcal{O}_i = \sum_k C_{i j k} \mathcal{O}_k \]
  which has closure and obeys the Jacobi identity. This algebra sets up the
  quantum kinematics.
  
  Further, a great simplification is afforded by a deep classical analogy.
  Corresponding to the classical canonical variables $\{ x_1 \ldots x_N, p_1
  \ldots p_N \}$ there exists a set of quantum variables such that
  \[ [x_i, p_j] \equiv x_i p_j - p_j x_i = i \hbar \{ x_i, p_j \}_{\tmop{PB}}
     = i \hbar \delta_{i j} \]
  where the $x_i$, $p_j$ in the third expression are classical variables and
  PB denotes the Poisson bracket. Most operators $\mathcal{O}_i$, hermitian and
  unitary, can be algebraically constructed out of the
  canonical set, with a few notable exceptions such as spin.
  
  Another important feature of quantum theory is that in Quantum Field Theory
  one also needs anti-commutator kinematic conditions. These have no classical
  analogue.
  
  \item {\tmstrong{Quantum dynamics.}}\label{pos:dynamics}
  
  a) In analogy with Classical Mechanics, there exists a distinguished
  hermitian operator, the Hamiltonian. In the Heisenberg picture dynamical
  evolution is expressed in terms of time dependent operators which can be
  observables or other operators
  \[ i \hbar \frac{d}{d t} \mathcal{O} (t) = [\mathcal{O} (t), H] \]
  In particular we have the analogues of Hamilton's equations of motion for the
  canonical variables,
  \[ i \hbar \frac{d}{d t} p (t) = [p (t), H] ; \qquad i \hbar \frac{d}{d t}
     x (t) = [x (t), H] \]
  from which the evolution equation for any operator on the phase space can be
  worked out using the canonical commutation rules.
  
  b) A convenient alternative is the Schr{\"o}dinger picture in which the
  state vector is time dependent, $| \Psi t \rangle \nobracket$. In this case
  the operators are not to be evolved in time, and the equation of motion for
  the state function is
  \[ i \hbar \frac{\partial}{\partial t} | \Psi t \rangle = H | \Psi t
     \rangle \]
  In practice we do not work with the abstract state $| \Psi t \rangle$, 
  but the ``wave function'' 
  $\Psi (x, t) \equiv \langle x | \Psi t \rangle$ and express 
  $H(x,p)\equiv H(x, -i\hbar d/dx)$. 
  
  c) A third very elegant and fruitful formulation of the dynamics is due to
  Dirac and Feynman, the Path Integral version. Usually the basis set $| x
  \rangle$ is treated as time independent. Now define a "moving basis" 
  (see \cite{DiracPQM}, Sec. 32) which we might call Dirac picture basis, 
  \[ | x t \rangle_D = e^{i H t / \hbar} | x \rangle \]
  Then the Path Integral formula gives the amplitude for going to $x_f$ at
  time $t_f$ given that the system was at $x_i$ at $t_i$ :
  \[ _D \langle x_f t_f | x_i t_i \rangle_D = \int \mathcal{D}x (t)
     \mathcal{D}p (t) \exp \left\{ \frac{i}{\hbar} \int_{t_i}^{t_f} d t (p
     \dot{x} - H) \right\} \]
  where the action integral in the exponent is on the phase space and the
  symbolic integration $\mathcal{D}x (t) \mathcal{D}p (t)$ is over all
  possible paths connecting $x_f $ and $x_i$. Then we can obtain $\Psi (x, t)$
  from $\Psi (x_i, t_i)$ for $t>t_i$ as 
  \[ 
  \Psi (x, t) = \int dx_i \ {_D \langle x t | x_i t_i \rangle_D} \Psi (x_i, t_i)
  \]
  
  \item {\tmstrong{Bosons and fermions.}}\label{pos:bosefermi}
  
  Another deep and non-classical feature of the quantum world is the existence
  of ``identical quanta''. These bits of nature have just a few attributes
  such as mass, spin and a few charges. In weakly coupled systems, the full
  multi-quanta Hilbert space can be constructed out of repeated tensor product
  of the one-quantum Hilbert space. This is called the Fock space. \
  
  The admissible multi-quanta states are only the symmetrised ones in an
  assembly of integer spin quanta, while the admissible states are only the
  anti-symmetrised ones for half-integer quanta. For the case of two quanta,
  with the states labeled by the values of the observable $\alpha$, these
  tensorial constructions are
  \[ | \Psi \rangle_B = \frac{1}{\sqrt{2}} \{ | \alpha^{(1)}_1 \rangle |
     \alpha_2^{(2)} \rangle + | \alpha^{(2)}_1 \rangle | \alpha_2^{(1)}
     \rangle \} \]
  \[ | \Psi \rangle_F = \frac{1}{\sqrt{2}} \{ | \alpha^{(1)}_1 \rangle |
     \alpha_2^{(2)} \rangle - | \alpha^{(2)}_1 \rangle | \alpha_2^{(1)}
     \rangle \} \]
  where subscripts $B$ and $F$ refer to Bose and Fermi respectively. 
\end{enumerateromancap}
This ends the list of the postulates incorporating the most essential rules. To repeat in a nutshell, the postulates refer to
\begin{enumerateromancap}
  \item Superposition principle and Hilbert space,
  
  \item Observables as hermitian operators,
  
  \item Change of basis as unitary operators,
  
  \item Observation as projection operator, 
  
  \item Kinematics as commutation rules,
  
  \item Dynamics via a special unitary operator, and
  
  \item Multi-quanta states via Bose-Einstein and Fermi-Dirac rules
\end{enumerateromancap}

\subsection{Remarks on the postulates}
\label{sec:remarks}
\begin{enumeratenumeric}
  \item The Hilbert space postulate captures two important things in one.
  Firstly there is superposition principle for the states, the deepest
  non-classical feature of the quantum world. Secondly we introduced the inner
  product on the space, as required for physical interpretation.
  
  \item Many cases of change of basis correspond to classical symmetry operations. 
  Wigner's theorem proves that such transformations are indeed
  represented by unitary or anti-unitary operators. Thus
Postulate \ref{pos:Changebasis} is not entirely an independent postulate, but 
it is important enough to be listed here.
  
  \item \label{point31-3} The issue of measurement has been a source of creative proposals and
  long standing debate among the finest minds. There are two aspects -- the
  outcome is probabilistic and there is no satisfactory description of the
  evolution from the unmeasured to the measured state. Here my first caveat is
  that a conscious observer may not be a key component of
  the measurement paradox. Scattering processes and spontaneous decay are directly 
  observed phenomena which capture most of the unsettling aspect of ``collapse 
  of the state'', with purely quantum evolution.
  \begin{enumerateroman}
    \item Intentional measurement has a lot of
    resemblance to scattering. In Rutherford type scattering we send in a
    stream of projectiles, which are momentum eigenstates. The expected out
    state obeys the symmetries of the scatterer, for example the azimuthal
    rotation symmetry along the direction of the incoming projectile. However, 
    a particular scattered particle
    can emerge only in one fixed direction. Thus while the evolution operator
    is unitary, the outcome for a single scattered particle is in a momentum eigenstate
    projected out from the evolved state.
        
    Only when a large number
    of the same projectiles with the same impact parameter is studied do we
    recover the azimuthal  symmetry. This can be considered to be 
    a ``collapse'' into the eigenvector representing that momentum value.
        Note that scattering goes on in locations remote from any conscious
    observer all over the Universe. The emergence in any one direction of the
    scattered particle is a generic event.
    
    \item Spontaneous decay is a similar phenomenon.  The emitted final state
    may be expected to obey the symmetries of the decaying parent ( for example 
    rotational symmetries of an atom or a molecule). And indeed
    this is so on the average. However any particular decay results in the 
    particle emerging
    only in a specific direction, and the symmetry can be recovered only
    through repeated experiments. Again, this process happens exactly thus,
    with no observer needed, though there could be one, light years away.
    
    It is interesting to note that this issue is implicit in Einstein's 1905 paper thus\cite{EinsteinMiraculous} :
    \begin{quotation}
According to the assumption considered here, in the propagation of a light ray 
emitted from a point source, the energy is not distributed continuously over 
ever increasing volumes of space, but consists of a finite number of energy 
quanta localised at points of space that move without dividing, and can be 
absorbed or generated only as complete units".
\end{quotation}
It is clear that in the emission of any individual quantum, the rotational 
symmetries of the source can not be respected. In turn, if the symmetry is 
to be recovered over a large number of observations, it should not be a 
surprise that the question of which direction is determined only
by a probabilistic law. In hindsight one may wonder why the bearer of so lucid
and profound an insight shied away from the collateral logical consequences.
  \end{enumerateroman}
  
  \item Returning to the broader issues of measurement, the problem seems to lie in the inability to characterise where the validity of classical paradigm ends and quantum regime begins. In some sense this is due to  the fact that $\hbar$ dimensionally involves space, time and mass and it is difficult to demarcate
  the transition between the two frameworks purely in terms of length or time or mass scales.
  Indeed Bohr's Correspondence Principle relies on largeness of quantum numbers to recover a classical description.
  
  In the Schr{\"o}dinger's Cat paradox the presumption is that all systems must rightly be considered as quantum.  In other words taking the observer to be entangled with the system being observed. Further, the cat box plus its observer can be a combined system being observed by another observer, thus demanding a combined description with indefinite recursion. This is the  "Wigner's Brother" paradox. 
  
  There are two possibilities for a resolution. Perhaps more delicate experiments will demand a more sophisticated formalism for their description. But if not, we only need improved semantics. A characteristic of macroscopic systems is that they are highly complex, say a Geiger counter or a bubble chamber, making it clear that they can be 
  simply   treated classically, thus at least avoid the recursion paradox. However the postulate as presented here has been verified
  in a variety of experiments directly or indirectly
  over the past century. While the paradox may persist, there is 
  no contradiction with the proposed postulate.\quad
  
  \item The Einstein-Podolski-Rosen paradox   epitomises another aspect that is
  counter-intuitive about measurement. 
  Suppose we have a two-electron state with net spin $=0$. Suppose an observer Alice at some remote point
  $\mathbf{x}$ makes an observation at time $t_1'$ observing only one of the electrons, and that in spin-up state,   destroying that electron in the process, then
  causality demands that at all subsequent times $t>t'_1$, the other observer Bob can only find one electron  and that in spin-down state.
  According to Special Relativity, this latter fact cannot in principle be
  known to Bob during the time $t'_1<t < t'_1+|\mathbf{x}|/c$. 
  In case Bob makes an observation during the time $t'_1<t <t'_1+ |\mathbf{x}|/c$ 
  they may legitimately ascribe the outcome to the weightage factor $1/\sqrt{2}$ for the spin-down state,   and nothing will go wrong. 
  The main lesson is that quantum states are intrinsically
  non-local and yet consistent with Special Relativity. The advanced framework
  of Relativistic Quantum Field Theory does not throw up any contradictions either.
  It does require the existence of anti-particles in order to preserve causality.
  More on the possible fallacy in EPR in the last section.
  
  \item We make a big leap of faith in assuming that all the observables and
  other operators can be expressed in terms of the canonical set. We propose
  that the $\mathcal{O}_i$ have the same algebraic dependence on the canonical
  variables as the corresponding classical variables. This is an immense
  simplification. But it requires a price to be paid
  \begin{itemizeminus}
    \item It results in operator ambiguity, which however is resolved by
    simple prescriptions.
    
    \item One may encounter observables with no classical analogue, e.g., spin
    
    \item For fermion fields one needs anti-commutators instead of
    commutators. There is no classical limit available for this operation in
    the normal sense.
  \end{itemizeminus}
  Fermion bilinears do have classical limits and one arranges to set up
  correct commutation relations between them with desirable classical
  limits\footnote{One might wonder whether this and the preceding point about
  spin have any mutual connection since fermions are prime examples of
  particles with spin. But integer spin particles also exist, and therefore
  the need to insert spin by hand will persist and is independent from the
  enigmatic quantisation of fermions.}.
  
  Despite all these exceptions, this principle must be viewed as of special
  significance. It suggests that the macroscopic canonical structure of
  Hamiltonian dynamics is firmly rooted in microscopic principles.
  
  \item Quantum kinematics can be equivalently expressed by the overlap between
  the bases labeled by canonical observables. This is given by the
  fundamental relation
  \[ \langle x| \nobracket p \rangle = \frac{1}{\sqrt{2 \pi \hbar}} e^{ipx /
     \hbar} \]
  This is the key building block of the Path Integral formula.
  
  \item Postulate \ref{pos:bosefermi} is most likely not a logically
  independent one. When Lorentz invariance is imposed on a quantum system, the
  ``spin-statistics theorem'' can be proved. We state this here as a law to
  highlight its importance. That is, that in Quantum Mechanics states are
  fundamental, not quanta. And to remind that ``quanta are not particles''.
  
  \item Quantum Field Theory provides the comprehensive framework of
  calculation in Fock Space based on weakly coupled quanta. However, the
  Quantum Field Theory framework has a far greater reach, in a wide variety of
  strongly coupled systems.
\end{enumeratenumeric}
\section{Is the Classical all reasonable?}\label{sec:classical}

The framework of Quantum Mechanics couched in the language of the Hilbert
space is at once forbidding to those unfamiliar with advanced Mathematics.
However there exists several quantum systems where this Hilbert space may be
no more forbidding than a finite dimensional complex vector space for a finite
number of spins. This is very promising because it may become possible to
teach simplified basics to even highschool students who may be able to
understand the technologies that are now emerging.

Despite this, the framework has become notorious for resorting to purely
abstract and therefore perhaps auxiliary constructs, and for being difficult to grasp
instinctively. As a counter to this, I take up here to argue that a few key
concepts of Classical framework are equally counter-intuitive, and it is more
the familiarity developed over several centuries that makes them so easily
acceptable.
\begin{enumeratenumeric}
  \item Instantaneous velocity : The most commonly held drawback of Quantum
  Mechanics is the loss of the detailed trajectory of the particle, specified
  by position and velocity at any given instant of time. It is however worth
  pondering how justified this expectation was in the first place. The notion
  that an object \tmtextit{is} at a place and \tmtextit{is moving} while at
  that place actually defies common sense. In a philosophical vain one is led
  to wonder how something can be {\tmem{at}} a place if it \tmtextit{is
  moving}. Newton solved this problem through the Calculus notion of limits.
  One defines
  \[ \tmmathbf{v}= \lim_{\Delta t \rightarrow 0}  \frac{\Delta
     \tmmathbf{x}}{\Delta t} \]
  The conceptual proposal however has never been verified empirically. No one,
  in Galileo's vain, took a meter stick and a stopwatch watch of unlimited
  precision and allowed the $\Delta \tmmathbf{x}$ and $\Delta t$ to physically
  go to zero. If they did they would discover Quantum phenomena. Thus there
  was no loss of paradise of unlimited precision, it was never there in the
  first place.
  
  To repeat Dirac's point of view quoted in the Introduction, $\hbar$ indeed
  sets a scale separating the Classical from the Quantum. In the limit the
  action function governing our moving particle reaches this level, the
  Newtonian paradise withers away and the Quantum emerges.
  
  This does draw our attention further to the carefully constructed notion of
  the continuum in Mathematics which buttresses Newton's conception.
  Development of science may have reached a point making it worth revisiting
  the tenets of the continuum. \ And in turn we need to develop mathematics in
  which quantum constructs take an intuitive centre stage.
  
  \item The abstract nature of the wave function is often contrasted with
  other physical quantities of immediate empirical meaning. The notions of
  electric and Magnetic fields are however worth re-inspecting. They are only
  a step away from being something palpable, since you need to hold a point
  charge or a small dipole in that region of space to immediately see their
  manifestation as field lines as visualised by Faraday. However when we
  propose that these fields oscillate and they carry energy, and they migrate
  over light years, in vacuum as Einstein taught us, then we may wonder
  ``what'' is actually propagating in the empirical sense. These fields
  therefore have a status comparable to that of the wave function.
  
  What is intriguing then is that according to the Sudarshan-Glauber theory,
  the quantum theory of light subsumes all its classical states without any
  change. Perhaps the validity of the conception of electromagnetic radiation
  is ultimately derived from the more fundamental quantum property.
  
  \item Classical conception is willing to live with point particles as well
  as continuum fields. However there is an internal contradiction in the
  Electrodynamics of a point charge. An oscillating point charge radiates, but
  the field of the charge cannot act upon itself. \ Thus the charge loses
  energy without any force slowing it down. Lorentz worked to give a finite
  size and an internal structure to the electron without success. If the field
  of a point charge is allowed to act upon itself we get an infinite answer.
  The issue occupied many stalwarts, Dirac, Wheeler and Feynman, and others.
  Its futility became clear after it was realised that the electron can be a
  point particle, yet obey quantum rules and never be localised to an ideal
  point. The questions in the revised version got transferred to those of a
  calculable framework of Quantum Electrodynamics, eventually addressed by
  renormalisation.
\end{enumeratenumeric}

\section{Exorcising the bogeys}\label{sec:bogeys}

A large number of professionals putting Quantum Mechanics successfully to use
remain unsure about its consistency and validity. To the extent these may only
be anxieties transmitted by the previous generation it is worth examining the
various enigmatic issues, and to finally banish the bogeys which can be
resolved in the light of the postulates formulated here.

To this end I go through a list of terms commonly used which continue to
reinforce a negative perception. It is my appeal that it is appropriate to
replace some of the description with better words and if this is not possible,
at least to stop using the terms in that form.
\begin{enumeratenumeric}
  \item {\tmstrong{The uncertainty principle}}. This essential novelty of
  quantum phenomena was gleaned by Heisenberg before arriving at full quantum
  mechanics. It holds a central role in our understanding of quantum systems.
  It provides a handy tool for making important estimates, also extended
  to energy-time uncertainty relation. However the term ``uncertain'' brings
  in serious negative connotation, which is worth avoiding.
  
  As all textbooks discuss \cite{SchiffQM}, this principle is a direct consequence of the
  superposition principle and the kinematics of canonically conjugate
  variables, postulates I and V.
  
  \item {\tmstrong{Wave function}}.\label{point5-2} The concept of the wave function is of
  great utility especially in visualising bound states and the electron
  probability density distribution in an atomic orbital. But the concept seems
  to get confused with waves on water or on stretched membranes.
  
  In Dirac notation, $\Psi (x) = \langle x| \Psi \rangle \nobracket$, or it is
  nothing but the component of the state along the basis vector $|x \rangle
  \nobracket$
  \[ | \Psi \rangle \nobracket = \sum_x |x \rangle \langle x| \Psi \rangle
     \equiv \sum_x C_x |x \rangle \nobracket \equiv \sum_x \Psi (x) |x \rangle
     \nobracket \]
  showing that it has the status of component $C_x$. However the wave function 
  can be obtained in any
  convenient basis such as the momentum basis $\Phi (p, t) \equiv \langle p |
  \Psi t \rangle \nobracket $, and convenient for bound states in spherically symmetric potentials,   $\Phi (l,m, t) \equiv \langle l,m | \Psi t \rangle \nobracket $.
  These can be interpreted as the
  corresponding \ ``components'' of the abstract state vector in that basis.\ 
  Indeed position space
  has a special status, more than any other observable. One then accords great
  primacy to this description, leading to mental pictures of catastrophic
  ``collapse'' of the wave function all over space.
  
  What about the fact that we have de Broglie wavelength associated with a
  particle? A free particle is in a momentum eigenstate, with some value $p$.
  Due to the existence of the fundamental constant $\hbar$, we can come up
  with a quantity of dimension of length, $\lambda = 2 \pi \hbar / p$. In
  hindsight, once the pilot waves are abandoned, there is nothing more to 
  the ``wave picture'' than this convenient
  re-expression of the momentum eigenvalue. The argument extends to a ``wave
  packet'' which is a linear superposition of momentum eigenstates.
  
  \item {\tmstrong{Wave particle duality}}. The term duality immediately
  suggests ambiguity of description. The classic experimental situation where
  wave particle duality can be demonstrated is the double slit experiment. If
  we can trace the electron as streaming through one of the two slits we call
  it the particle manifestation, if we forgo the knowledge of which slit it
  traveled through we obtain a fringe pattern characteristic of wave
  phenomena.
  
  Let us be clear that even in the ``wave'' avatar, the electron is going to
  be actually measured as a particle, by a click counter. The number of clicks
  will now be distributed in the fringe pattern. Further, the wave is only a
  mental construct obtained by relabeling momentum eigenstates in terms of
  ``wavelengths'' $\lambda$ as discussed in Wave Function topic above.
  
  But don't we still face the paradoxical situation? Not if we properly digest
  postulate V, the canonical commutation relation. Effectively we are switching between position observable ( which slit) and momentum observable (free streaming to
  interfere) and the what we observe is easily deduced from Kinematics. 
    We can summarise the
  situation by saying that the double slit experiment will remain a key part
  of pedagogical discussion as included in Sec. 2, being the most direct
  manifestation of quantum phenomena. But it needs formalisation as Postulate
  V and accepted as a part of the basic tenets rather than be mystified.
  
  \item {\tmstrong{Collapse of the wavefunction}}. We have discussed above the
  convenience but also the secret of the wavefunction. Due to primacy of space
  variable, our postulate IV gets construed as ``collapse of the wave
  function''. As we remarked in \ref{point31-3}, Sec \ref{sec:remarks}, the enigmatic situation with
  respect to probabilistic outcomes will remain or get resolved with more
  experiments. In the meantime we need to banish this term and live with this
  postulate.
  
  \item {\tmstrong{Indistinguishability}}. We tell students about
  indistinguishable quanta. The fallacy is that there was nothing to
  distinguish to begin with. The term is only a hang over of the classical
  ideas about billiard balls. Two of them machined with great care may be
  impossible to distinguish. It is with respect to this classical expectation
  that we dub the quanta to fall into the same descriptive category.
  
  More carefully stated, in weakly coupled systems number is an observable
  quantity and these are eigenstates of that observable with value 2 or more.
  For photons the number is not conserved, but it is a valid observable. Note
  that regardless of how many quanta it has, the state is normalised to unity.
  
  We can trace two sources of the misnomer. One is that the many-quanta states
  in weakly coupled systems are most conveniently obtained as tensor products
  of 1-particle or 1-quantum states. This accords an unjustified primacy to
  the 1-quantum state, and one thinks of the purely mathematical construct as
  the physical one of two or more quanta being brought together, whereas it 
  is a single state in the Hilbert space.
  
  Another is a more empirical source. A state of two quanta reduces to a state
  of one quantum if one of them is absorbed or undergoes observation. And \
  ``which'' quantum got observed and ``which'' remained is a meaningless
  question. But after the observation of one of them we have two distinct
  ones. It is with respect to this putative final state that we call that
  system to be contain indistinguishable quanta. \ 
  
  \item {\tmstrong{Statistics}}. It is common to refer to the Bose-Einstein and
  Fermi-Dirac prescriptions to be ``statistics''. This may leave the student
  with the idea that some randomness or incompleteness is present and we are
  sampling partial information, as would be done in macroscopic setting.
  
  The prescriptions are only providing the correct enumeration of states as
  tensor products of 1-quantum states, with nothing statistical about the
  prescription. The first manifestation of these uncanny rules was deduced in
  the context of thermodynamics of quantum gases. It is perhaps a left over
  terminology from that.
  
  \item {\tmstrong{Entanglement.}} A state of several quanta is merely the
  correctly identified basis state of the weakly coupled system. \ There are
  really no entities undergoing ``tangling'' with each other. Only if we refer
  to a post facto state of quanta subjected to observation that we can say
  there were several different entities which could have been entangled. Per
  se there are just many-quanta states as per B-E or F-D rules with nothing
  tangled up inside them. Correlations is a less drastic term, though still
  with respect to post observation states.
  
  \item {\tmstrong{The EPR paradox}}. This is now a composite of several of
  the issues already described. The conceptual error is to think that there
  are ``two'' systems. It is actually {\tmem{one }}quantum state normalised to
  unity, with two {\tmem{quanta}}, the latter number being the eigenstate of
  the number operator. When evolved, it can have configuration space component
  (in the sense of point \ref{point5-2}, Wavefunction, above) which is highly non-local. \
  Observation process breaks the quantum system, with the number operator
  eigenvalue having reduced to one in the remainder state. The remaining
  system needs to be freshly normalised to unity. The uncanny situation should 
  not be held up as mysterious or unsatisfactory. It  only needs to be referred to postulate IV on Observation.
\end{enumeratenumeric}

\section{Conclusion}\label{sec:conclusion}
We have summarised an essential core of the well known principles of Quantum Mechanics itemised as seven postulates. They are not logically independent in the axiomatic sense 
but are of primary importance to be listed independently. It is necessary that students develop direct intuition about raw quantum facts, and understand them as the sources of 
the mathematical laws. As such we discussed relevant basic experiments. I have tried to
dispel the negative perception in which a variety of quantum phenomena are held. 
After one century of success of the theory one must jettison the baggage of the past 
and attempt to come up with better terminology, or at least avoid using its negative 
contents. I have tried to argue that if one properly digests the truth of the 
postulates, most of the negative connotations and the 
sense of spooky mystery should get dispelled. Finally and most importantly, I have 
pointed out the fallacy of instantaneous velocity which gives rise to the illusion 
of perfect predictive power of classical mechanics. Indeed with hindsight one may 
note that the continuum  is a purely logical construct, a description not borne out 
by the material media once assumed to be continua. Functional analysis nevertheless 
continues to be the bulwark of Quantum Mechanics, albeit after due tweaking. Perhaps 
there are more subtle formulations of the continuum that make contact with the 
reality of fundamental quantum phenomena more easily.

\end{document}